\documentstyle[12pt]{article}
\newenvironment{namelist}[1]{%
\begin{list}{}
{

\settowidth{\labelwidth}{#1}
\setlength{\leftmargin}{1.05\labelwidth}
}
}{%
\end{list}}
\newcommand{\dis}{\displaystyle}
\begin{document}
\def\qed{~\vrule height6pt width4pt depth0pt\medskip}
\setlength{\baselineskip}{0.34in}
\setlength{\jot}{0.2in}

\begin{center}
{\large \bf
On the passibility of a short-lived free Dirac particle
}\\

\vspace*{0.5cm}
Lu Lin
\\
Department of Electrophysics\\
National Chiao Tung University\\
Hsinchu, Taiwan \\
Republic of China\\
\end{center}

\vspace*{2cm}

\noindent
{\bf ABSTRACT}~~\\

A time-decaying Dirac equation is suggested.  For a free particle
with rest mass about 0.5GeV, the live-time is about $10^{-25}$ second
in the rest frame.\\

\vspace*{2cm}

\noindent
PACS numbers: 03.56.Pm, 12.90.+b, 14.80.-j
\newpage
In two recent articles [1], the author has discussed some properties
concerning a free Dirac electron.  Based on Dirac equation, a possibility
of a complex 4-dimensional spac-time manifold was suggested, which
leads to a function for describing a free particle should be a free wave.
Furthermore since there are couplings between the spatial variables
and the internal variables $(\alpha _{i}p_{i})$ [2] in the Hamiltonian,
the generalized coordinates become mixtures of spatial coordinates and the
internal coordinates.
In fact, Dirac [2] calculated that $x_{1}=a_{1}+c^{2}p_{1}H^{-1}t
+\dis\frac{i}{2}c\hbar (\alpha_{1}-cp_{1}H^{-1})H^{-1}.$
 As a consequence the velocities of the genrealized
coordinates $\dot{x}_{i}'s$ are $c\alpha_{i}$.  However, the velocities
of the spatial coordinates $v_{i}=p_{i}H^{-1}(i=1,2,3)$  which
commute with $H$ and can be simultaneously measured with $H$.  That is,
there two kinds of velocities for a free Dirac particle.  In this
note, it is suggested that a short-lived free Dirac particle may exist
which may be described by a time decaying Dirac equation.\\

Given a Hamiltonian
\begin{eqnarray}
H=c(p_{x}^{2}+p_{y}^{2}+p_{z}^{2}+m^{2}c^{2})^{1/2}.
\end{eqnarray}

\noindent
Let us define a Dirac process as the following: Introduce the $\alpha '_{i}s$
and $\beta $ together with their relations as defined in ref.[2].  Then
take the square root of the right-hand-side of equation (1),
we obtain the Dirac Hamiltonian
\begin{eqnarray}
H=c\mbox{\boldmath $\alpha $}\cdot \mbox{\boldmath $p$}+mc^{2}\beta .
\end{eqnarray}

\noindent
In this way, the internal variables are introduced.
 Meanwhile, interactions between the spectial
variables and the internal variables are also introduced.  The
generalized coordinales are created and become mixtures of the spatial
varibles and the internal variables.  Also, two kinds of velocities are
produced.\\

In classical machanics, a Hamiltonian $H(p,q)$ is defined for a free
particle as
\begin{eqnarray}
H(p,q)=\sum p_{i}\dot{q}_{i}-L(q_{i},\dot{q}_{i}),
\end{eqnarray}

\noindent
where the $q_{i}'s, \dot{q}_{i}'s$ and the $p_{i}'s$ are the generalized
coordinates, generalized velocities and genrealized momenta.
The $\dot{q}_{i}'s$ are supposed to be eliminated by some available
relations.  For a relativistic free particle, $L$ is given as [3]
\begin{eqnarray}
L=-mc^{2}(1-\frac{1}{c^{2}}\sum \dot{q}_{i}^{2})^{1/2}.
\end{eqnarray}

\noindent
If we substitude equation (4) into equation (3) and elimmate the
$\dot{q}_{i}s$ by $\dot{q}_{i}=p_{i}H^{-1}$, we will have equation (1).
Then we can perform a Dirac process to obtain Dirac equation.  However,
if we perform the Dirac process one step before we get to equation (1).  That
is, we perform the Dirac process for the function $L$ of equation (4) and
substitute into equation (3),
then we have
\begin{eqnarray}
H(p,q)=\sum p_{i}\dot{q}_{i}+mc^{2}(\beta \pm \frac{i}{c}\sum
\alpha_{i}\dot{q}_{i}),
\end{eqnarray}

\noindent
where the $\pm $ sign is determined by the condition for a finite
solution with respect to the forward or backword propergating wave.
Now we have a Dirac type of Hamiltonian where the $\dot{q}_{i}'s$
are to be eliminated by suitable relations.  If we tak
$\dot{q}_{i}=p_{i}H^{-1}$, it is straightforward to show that
equation (5) reduces to equation (2) by noticing that,
$H^{2}-c^{2}p^{2}=(mc^{2})^{2}$, and $(-i\beta \alpha_{i}'s, \beta )$
is equivalent to $(\alpha_{i}'s, \beta )$.  We are aware that the
Hamiltonian (5) is supposed to describe a free Dirac particle, if this
is correct, then a Dirac free particle has two kinds of velocities,
velocities of the generalized coordinates and those of the spatial
coordinates.  Therefore we can also take $\dot{q}_{i}=c\alpha_{i}$ in
equation (5).  Thus we obtain
\begin{eqnarray}
H=\sum c\alpha_{i}p_{i}+mc^{2}\beta \pm 3imc^{2}.
\end{eqnarray}

\noindent
This Hamiltonian will result with a solution having a couplex energy
that represents a particle with a life-time about $\hbar /(3mc^{2})$.
For a particle with rest mass about 0.5 GeV, the life-time is about
$10^{-25}$ second in the rest frame.  In the lab-frame, its life-time
can be determined by relativity.\\

We conclude that a Lagrange-Hamilton formalism may lead to two kinds
of Dirac particles, one kind of them are stable particles, the other
unstable.  Finally, one may wish to ask whether one of those un-discovered
particles, such as a quark, may belong to the second kind of particles.
If an unstable particle of this kind would be created from a source,
it can only live for such a short time, then decay immediately.
 If there is no available exit channel,
it
must go back to the source again.\\

\noindent
References
\begin{namelist}{[1]}
\item [{[1]}]
Lu Lin, ``On the possibility of a complex 4-dimensional space-time
manifold'', Gen. Phys./9804016; Lu Lin, ``The velocity and angular
momentum of a free Dirac electron'', Gen. Phys./9804024.
\item [{[2]}]
P.A.M. Dirac, The principles of quantum mechnaics, fourth ed. 1958,
Oxford Univ. Press, London.
\item [{[3]}]
L.D. Landau and E.M. Lifshitz, The classical theory of fields,
 translated from Russian by
M. Hamermesh, 2nd Printing 1959, Addison-Wesley series in Advancd Physics.
\end{namelist}

\end{document}